\documentclass[12pt]{iopart}
\usepackage{graphicx}

\begin{document}

\title{Role of density fluctuations in the relaxation of
  random dislocation systems}

\author{Ferenc F Csikor$^{1,2}$, Michael Zaiser$^{2}$,
  P\'eter Dus\'an Isp\'anovity$^{1}$ and Istv\'an Groma $^{1}$}

\address{$^{1}$ Department of Materials Physics, E\"otv\"os
  Lor\'and University, P\'azm\'any P\'eter s\'et\'any 1/a, H-1117
  Budapest, Hungary}
\address{$^{2}$ Centre for Materials Science and Engineering, The
  University of Edinburgh, The King's Buildings, Edinburgh EH9~3JL,
  United Kingdom}
\eads{\mailto{csikor@metal.elte.hu}, \mailto{M.Zaiser@ed.ac.uk},
  \mailto{ispanovity@metal.elte.hu}, \mailto{groma@metal.elte.hu}}

\begin{abstract}
  We study the relaxation dynamics of systems of straight, parallel
  crystal dislocations, starting from initially random and
  uncorrelated positions of the individual dislocations. A scaling
  model of the relaxation process is constructed by considering the
  gradual extinction of the initial density fluctuations present in
  the system. The model is validated by ensemble simulations of the
  discrete dynamics of dislocations. Convincing agreement is found
  for systems of edge dislocations in single slip irrespective of the
  net Burgers vector of the dislocation system. It is also
  demonstrated that the model does not work in multiple slip
  geometries.
\end{abstract}

\pacs{
  61.72.Lk, % Defects and impurities in crystals; microstructure --
  % Linear defects: dislocations, disclinations
  05.40.-a, % Statistical physics, thermodynamics and nonlinear
  % dynamical systems -- Fluctuation phenomena, random
  % processes, noise and Brownian motion
  45.50.Jf, % Dynamics and kinematics of a particle and a system of
  % particles -- Few- and many-body systems
  62.20.fq  % Mechanical and acoustical properties of condensed matter
  % -- Mechanical properties of solids -- Plasticity and
  % superplasticity
}

\noindent{\it Keywords\/}: defects (theory), fluctuations (theory),
plasticity (theory)

\maketitle

\section{Introduction}
\label{sec:introduction}

The creation and motion of large numbers of crystal lattice
dislocations is the most fundamental feature of crystal
plasticity. During the last half century, the physical properties of
individual dislocations and their interactions with localised
obstacles have been studied extensively. On the other hand, the
complex collective dynamics of strongly interacting
many-dislocation systems is still far from being
understood. Fortunately, everyday plastic deformation processes
very often proceed orders of magnitude slower than the typical
relaxation times of the underlying dislocation system. These
conditions
often permit to study the problem in a quasistatic approximation
\cite{zaiser01b,groma03}. Beyond the quasistatic limit, however, much
less work has been devoted to studying the dynamics of collective
dislocation motions which lead to the formation of metastable
configurations, and to transitions between such configurations in
driven dislocation systems. However, such collective motions are
crucial for understanding rapid dislocation processes which not only
occur in shock loading but, in the form of dislocation avalanches, are
a generic feature of the dynamics of driven dislocation systems
\cite{zaiser06}.

The first studies of dynamic relaxation processes in dislocation
systems were performed by Miguel et al.\ with the protocol of applying
a constant external shear stress to well relaxed dislocation
configurations \cite{miguel02,miguel05,miguel07}. The ensuing creep
relaxation was numerically shown to follow Andrade's law stemming from
the underlying intermittent and correlated motion of dislocation
structures. The connection between the mesoscopic and macroscopic
features of the process was, however, not analysed in detail.

Another direction was taken by the present authors who conducted
systematic studies of the relaxation dynamics of initially random
configurations of straight dislocations. This is an important issue
since the elastic energy density ${\cal E}$ of a random dislocation
system of density $\rho$ is known to diverge with the logarithm of
system size $L$, ${\cal E} \propto \rho \ln(L/b)$ \cite{wilkens67}
where $b$ is the modulus of the dislocation Burgers vector. In a
well-relaxed dislocation arrangement, on the other hand, the same
quantity scales like ${\cal E} \propto \rho \ln(1/(b\sqrt{\rho}))$,
i.e., the screening length corresponds to the mean dislocation spacing
\cite{wilkens69,zaiser01,zaiser02}. As the mean square stress is
proportional to the elastic energy density, this screening also
removes a logarithmic divergence of the width of the internal stress
probability distribution \cite{csikor04}, and of the X-ray line width
\cite{zaiser02,csikor04,krivoglaz69}. Numerical experience showed
that, at least in single slip geometries, the relaxation processes
that lead to screened dislocation arrangements exhibit slow, power law
characteristics for quantities such as the elastic energy or the
average dislocation velocity \cite{csikor05}. A model was proposed
which relates the power-law relaxation dynamics to the gradual
extinction of initial dislocation density fluctuations
\cite{csikor06}. The present paper presents a comprehensive numerical
investigation which allows to check in detail the model predictions
and complements the earlier work by extending the investigation to
multiple slip geometries and to dislocation systems of non-zero net
Burgers vector, and by studying the influence of an external driving
stress on the relaxation process.

The paper is organised as follows. In \sref{sec:simulation} the
problem is defined and technical details of the simulations are
presented. \Sref{sec:theory} unfolds a scaling model of the relaxation
process from a chemical analogy and uses this model to predict the
evolution of simulation measurables. \Sref{sec:comparison} then gives
a detailed comparison between model predictions and numerical
results. The results are discussed and conclusions are drawn in
\sref{sec:conclusions}. An auxiliary calculation of the elastic energy
of a random dislocation wall is presented in the Appendix.

\section{Problem definition and simulations}
\label{sec:simulation}

\subsection{Equations of motion}
\label{sec:formulation}

Consider a system of $N$ straight edge dislocations running parallel
to the $z$ axis of a Cartesian coordinate system. Let all dislocations
have a common Burgers vector pointing along the $x$ axis (a so-called
single slip geometry), $\bi{b}_{i} = s_{i} b \bi{e}_{\mathrm{x}}$,
where $s_{i} = \pm 1$ is the sign of the $i$th dislocation. Assuming
overdamped glide motion with a dislocation velocity $v$ that is
proportional to the local resolved shear stress, and zero dislocation
mobility in the climb direction, the equation of motion of dislocation
$i$ piercing the $xy$ plane at $\bi{r}_{i} = (x_{i}, y_{i})$ can be
written as
\begin{equation}
  \label{eq:eqmdim}
  \dot{x}_{i} =
  \chi b s_{i} \left[ \sum_{j \neq i}
    s_{j} \tau_{\mathrm{ind}}(\bi{r}_{i} - \bi{r}_{j})
    + \tau_{\mathrm{ext}} \right],
  \qquad \tau_{\mathrm{ind}}(\bi{r}) =
  G b \frac{x (x^{2}-y^{2})}{(x^{2}+y^{2})^{2}},
\end{equation}
where $\chi$ denotes the dislocation glide mobility, $G = \mu / [2 \pi
(1 - \nu)]$ where $\mu$ is the shear modulus and
$\nu$ is Poisson's ratio of the embedding isotropic crystal,
$\tau_{\mathrm{ind}}(\bi{r})$ denotes the resolved shear stress
field induced by a positive dislocation located at the origin
\cite{hirthlothe82}, and $\tau_{\mathrm{ext}}$ is a constant
externally applied resolved shear stress.

It is useful to introduce natural coordinates at this point which will
be denoted by an apostrophe ($'$) in the following. Measuring length
in units of the average dislocation--dislocation distance
$\rho^{-1/2}$ (where $\rho$ denotes the total dislocation density of
dislocations including both signs and, in multiple slip geometries,
including all slip systems), stress $\tau$ in units of
$Gb\sqrt{\rho}$, and plastic strain $\gamma$ in units of
$b\sqrt{\rho}$ leads to the relations
\begin{equation}
  \label{eq:dimless}
  \eqalign{
    &x' = x \sqrt{\rho}, \qquad t' = t \rho \chi G b^{2}, \qquad
    \tau' = \tau / (G b \sqrt{\rho}), \cr
    &v' = v / (\sqrt{\rho} \chi G b^{2}), \qquad
    \gamma' = \gamma / (b\sqrt{\rho}), \qquad
    \Delta E' = \Delta E / (Gb^2),
  }
\end{equation}
where $\Delta E$ is the elastic energy difference between two states
of the system (energy per unit dislocation length). In natural
coordinates \eref{eq:eqmdim} takes the form
\begin{equation}
  \label{eq:eqm}
  \eqalign{
    &\dot{x'}_{i} = 
    s_{i} \left[ \sum_{j \neq i}
      s_{j} \tau'_{\mathrm{ind}}(\bi{r}'_{i} - \bi{r}'_{j})
      + \tau'_{\mathrm{ext}} \right], \cr
    &\tau'_{\mathrm{ind}}(\bi{r}') =
    \frac{x' (x'^{2}-y'^{2})}{(x'^{2}+y'^{2})^{2}} =
    \frac{\cos(\varphi) \cos(2\varphi)}{r'},
  }
\end{equation}
where $\varphi$ denotes the angle between the $x$ axis and $\bi{r}'$.

\subsection{Simulation details}
\label{sec:simdetails}

To study dislocation relaxation, a large number of discrete
dislocation dynamics simulations have been performed. Equations of
motion \eref{eq:eqm} were solved with the 4.5th order
Runge--Kutta--Fehlberg method. Periodic boundary conditions were
applied to a square simulation area with edges parallel to the slip
planes, following the method used in \cite{bako06}.

To avoid overly small timesteps during the final stages of approach of
narrow dipoles (pairs of dislocations of opposite signs), a small
number of extremely narrow dipoles were excluded from the solution of
\eref{eq:eqm} and forced to move as if they were isolated from the
rest of the system. This is justified as the far-field stresses of the
dislocations in a narrow dipole cancel, while their pair interaction
diverges when the dislocation--dislocation distance approaches
zero. As a consequence, the dynamics of narrow dipoles is effectively
uncoupled from the rest of the dislocation system. We do not allow for
annihilation of narrow dipoles, which is a process that is governed by
the atomistics of the dislocation cores. This implies that we consider 
dislocation spacings to be large in comparison with the dipole annihilation
distance, which is believed to be of the order of one nanometer. As typical 
dislocation densities in highly dislocated crystals are of the order of 
$10^{14}$~m$^{-2}$, i.e.\ the average dislocation spacings are 
of the order of a hundred nanometers, this is not a severe restriction.

A first set of simulations was started from random configurations of
equal numbers of positive and negative dislocations. The number of
simulated dislocations $N$ (which defines the system size
$L_{\mathrm{s}}'$ since $N = L_{\mathrm{s}}'^{2}$) varied between $16$
and $128$. It is well known that the flow stress of single-glide
dislocation systems is around $0.1$ in natural units
\cite{miguel02}. To allow the dislocation systems to reach mechanical
equilibrium at the end of the relaxation, we restricted the applied
external stresses to levels below the flow stress, using stresses
between $0$ and $0.088$ natural units.

\begin{figure}
  \centering
  \includegraphics{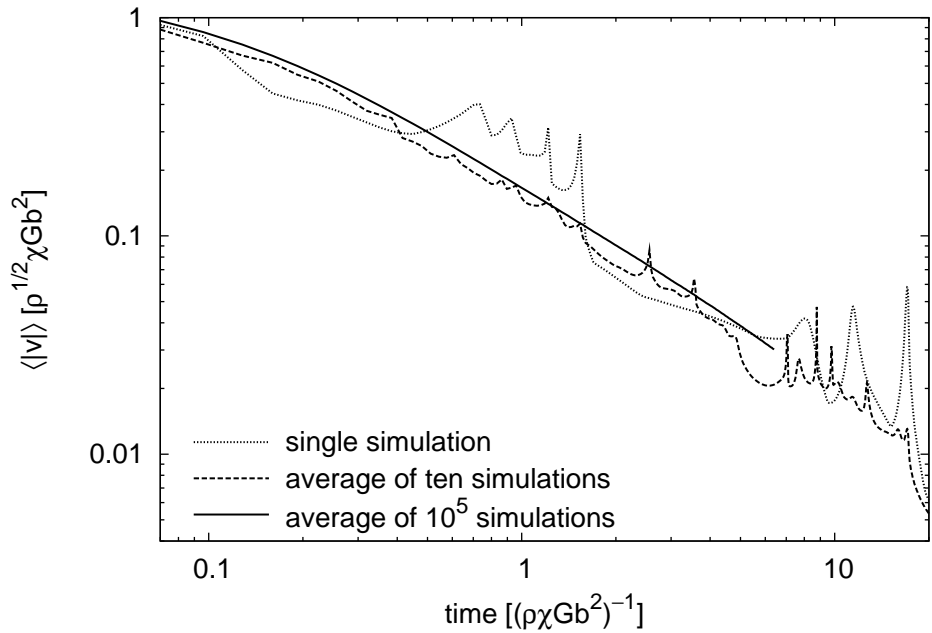}
  \caption{Evolution of the average dislocation velocity for different
    ensemble sizes.}
  \label{fig:averaging}
\end{figure}

As seen in \fref{fig:averaging}, individual simulations showed strong
avalanche-like activity during relaxation, as previously observed in
\cite{miguel05}. To reveal scaling properties of the relaxation
process, the evolution of global parameters such as stored energy,
mean absolute dislocation velocity, mean square velocity, and mean
strain rate was averaged over $3600$ to $10^{5}$ simulations starting
from different random initial configurations. This ensemble averaging
resulted in smooth ensemble averaged graphs as seen in
\fref{fig:averaging}.

In addition to the relaxation of `neutral' arrangements of
dislocations moving on a single slip system aligned parallel to the
edge of a square simulation area, we considered two variants of this
basic setting (these simulations were only performed at zero external
stress): i) To elucidate the influence of different implementations of
the boundary conditions, we performed single slip simulations with the
simulation box oriented at an angle $\varphi$ to the slip planes. This
does not affect the short-range dislocation--dislocation interactions
but modifies the stress field created by the periodic images. ii) To
study the influence of a net Burgers vector on the relaxation process,
we investigated the limiting case of fully polarised dislocation
systems (dislocations of one sign only) of sizes $16 \le N \le 128$.

To investigate the differences between single and multiple slip
geometries, we also performed simulations in which dislocations of
multiple slip systems were present. In these simulations we considered
sets of equally populated slip systems, each containing edge
dislocations with a zero net Burgers vector that were initially
distributed at random. The methodology of the multiple slip
simulations was identical to the single slip simulations described
above, with the following differences: i) For each dislocation in the
system, the complete elastic stress tensor was computed, again
assuming periodic boundary conditions in a square simulation area and
using the method of \cite{bako06}. The forces acting on the
dislocations were calculated from the stress tensor components using
the Peach--Koehler formula \cite{hirthlothe82}. ii) In addition to
narrow dipoles, which were treated in the same way as described above,
pairs of attracting dislocations on intersecting slip planes needed to
be treated separately. Such dislocation pairs react with each other
and form a reaction product (`dislocation lock') with Burgers vector
equal to the net Burgers vector of the constituent dislocations. In
real crystals, both the mobility of dislocation locks and the stress
required for separating them into the constituent dislocations depend
on their atomic core structure. For simplicity, we assumed all
dislocation locks to be immobile and to possess infinite separation
stress. New dislocations joining an existing lock were assumed to
annihilate with the constituent dislocation of the opposite Burgers
vector if such a dislocation was present. If a dislocation lock
converted to an ordinary dislocation through such an annihilation
event, it was assumed to become mobile again. To avoid overly small
timesteps during the final stages of dislocation lock formation or
reaction with a new dislocation (again due to diverging elastic
interactions between the constituent dislocations) the moving
constituent dislocations were pinned when their relative distance
decreased below a small predefined reaction radius which mimics the
core extension of the dislocation lock. For simplicity, the stress
field of a lock was calculated as the superposition of the stress
fields of the pinned constituent dislocations. Although the positions
of the constituent dislocations are scattered over a small region of
the size of the reaction radius, their net stress field is a good
approximation of the stress field of the lock at larger
distances. Finally, we note that this stress field is of long range
character since the net Burgers vector of a dislocation lock is not
zero.

\section{Theory}
\label{sec:theory}

In this section a simple scaling argument is used to establish the
asymptotic kinetics of a bimolecular combination reaction. We then
adapt the fundamental ideas behind this argument to the relaxation of
dislocation systems as described in the previous section. Based on the
adapted model, predictions are made for the evolution of several
physical quantities which can be directly obtained from the
simulations described in \sref{sec:simulation}. These predictions are
compared in detail to the numerical results in \sref{sec:comparison}.

\subsection{Role of density fluctuations in bimolecular reaction
  kinetics}
\label{sec:kinetics}

Consider the direct combination reaction
\begin{equation}
  \label{eq:reaction}
  \mathrm{A} + \mathrm{B} \rightarrow \mathrm{C}
\end{equation}
starting from a random, balanced configuration of the two reactants
\begin{equation}
  \label{eq:initial}
  \bar{c}_{\mathrm{A}}^{0} = \bar{c}_{\mathrm{B}}^{0},
\end{equation}
where $\bar{c}$ denotes the spatial average of the concentration field
$c(\bi{r})$. Note that the initial concentrations of molecules
$\mathrm{A}$ and $\mathrm{B}$ are equal only in an average sense
because of the thermally induced random positions of the individual
molecules. These concentration fluctuations need to be taken into
account when modelling the reaction kinetics. For situations where
long-distance transport of particles occurs by free Brownian motion, a
simple scaling argument which captures the essential physics was given
by Ovchinnikov \cite{ovchinnikov78} (which the reader is advised to
consult for further details). Broadly speaking, this involves the
subsequent dominance of two consecutive mechanisms directly
corresponding to the different length scales inherent in the system:
i) the reaction of those molecules that do not need long distance
motion to find a reaction partner, followed by ii) the long range
Brownian motion and subsequent reaction of excess reactants remaining
as a result of initial concentration fluctuations.

\subsubsection{Stage 1: Reaction controlled kinetics.}
\label{sec:secondorder}

As discussed in \cite{ovchinnikov78}, the first stage of reaction
\eref{eq:reaction} is controlled by a bimolecular reaction rate $k$
which depends on the short-range interactions between the reaction
partners. This stage can be discussed in the classical approximation
\begin{equation}
  \label{eq:rate1}
  \dot{\bar{c}}_{\mathrm{A}} =
  \dot{\bar{c}}_{\mathrm{B}} =
  -k \bar{c}_{\mathrm{A}} \bar{c}_{\mathrm{B}}.
\end{equation}
Solving \eref{eq:rate1} for initial conditions \eref{eq:initial}
yields
\begin{eqnarray}
  \label{eq:solution1}
  \eqalign {
    &\bar{c}_{\mathrm{A}} = \bar{c}_{\mathrm{B}} =
    \frac{\bar{c}_{\mathrm{A}}^{0}}{1 + k \bar{c}_{\mathrm{A}}^{0} t},
    \qquad t \ge 0, \cr
    &\bar{c}_{\mathrm{A}} = \bar{c}_{\mathrm{B}} = (kt)^{-1}, \qquad
    t \gg (k \bar{c}_{\mathrm{A}}^{0})^{-1}.
  }
\end{eqnarray}

\subsubsection{Stage 2: Diffusion controlled kinetics.}
\label{sec:diffusion}

Because of thermal concentration fluctuations, not all of the
molecules $\mathrm{A}$ and $\mathrm{B}$ can be consumed during  stage
1. Spatial fluctuations of the reactant concentrations lead to local
excess of molecules of one type. As these excess molecules cannot find
a local partner, they need to migrate via long range Brownian motion,
leading to a second kinetic stage controlled by long-range diffusion.

To characterise the concentration fluctuations in question we observe
that, on scales larger than the range of the `contact interactions'
which govern the first stage of the reaction kinetics, the positions
of molecules are statistically independent. As a consequence, the
initial numbers of molecules $\mathrm{A}$ or $\mathrm{B}$ in a
sufficiently large volume $V$ are Poisson distributed. This implies
that the mean numbers of excess molecules fulfil the relations
\cite{ovchinnikov78}
\begin{equation}
  \label{eq:initialexcess}
  \eqalign{
    &\left( \bar{N}_{\mathrm{A, excess}}^{0, V}\right)^{2} = V^2
    \langle (c_{\mathrm{A}}(\bi{r}) -
    \bar{c}_{\mathrm{A}}^{0})^{2} \rangle =
    V \bar{c}_{\mathrm{A}}^{0}, \cr
    &\left( \bar{N}_{\mathrm{B, excess}}^{0, V}\right)^{2} =
    V^2 \langle (c_{\mathrm{B}}(\bi{r}) -
    \bar{c}_{\mathrm{B}}^{0})^{2} \rangle =
    V \bar{c}_{\mathrm{B}}^{0},
  }
\end{equation}
where $\langle \rangle$ denotes spatial averaging over a large
statistically homogeneous system or, equivalently, ensemble averaging
over a large number of statistically independent and equivalent
realizations.

To further discuss the reaction kinetics during stage 2, we confine
ourselves to the limiting case of an infinite bimolecular reaction
rate $k \to \infty$. This choice does not affect the generality of the
discussion, it only affects the moment of the crossover between stage
1 and stage 2 kinetics. We introduce the concentration difference
$z(\bi{r})$ as
\begin{equation}
  \label{eq:defz}
  z(\bi{r}) = c_{\mathrm{A}}(\bi{r}) - c_{\mathrm{B}}(\bi{r}).
\end{equation}
In a hypothetical initial state before the reaction has been `switched
on' at $t=0$, $c_{\mathrm{A}}(\bi{r})$ and $c_{\mathrm{B}}(\bi{r})$
are statistically independent and $z(\bi{r})$ has the initial
statistical properties
\cite{ovchinnikov78}
\begin{equation}
  \label{eq:zrel}
  \eqalign{
    &\langle z(\bi{r}) \rangle =
    \langle c_{\mathrm{A}}(\bi{r}) - c_{\mathrm{B}}(\bi{r}) \rangle =
    0, \cr
    &\langle z(\bi{r})^{2} \rangle =
    \langle (c_{\mathrm{A}}(\bi{r}) - c_{\mathrm{B}}(\bi{r}))^{2}
    \rangle = 2 \bar{c}_{\mathrm{A}}^{0} / V =
    2 \bar{c}_{\mathrm{B}}^{0} / V.
  }
\end{equation}
To proceed, we note that for $k \to \infty$ molecules $\mathrm{A}$ and
$\mathrm{B}$ cannot coexist for $t>0$. Therefore, $z(\bi{r})$ gives a
complete characterisation of the concentration map for $t > 0$: for
$z(\bi{r}) > 0$, $c_{\mathrm{A}}(\bi{r}) = z(\bi{r})$ and
$c_{\mathrm{B}}(\bi{r}) = 0$ and for $z(\bi{r}) < 0$,
$c_{\mathrm{A}}(\bi{r}) = 0$ and
$c_{\mathrm{B}}(\bi{r}) = - z(\bi{r})$. Because there exists no other
physical length scale in the system, the size of the
regions characterised by $z > 0$ and $z < 0$ is determined by the
diffusion length $L(t) = \sqrt{Dt}$ (for simplicity, the same
diffusion constant $D$ is assumed for both kinds of reactants).

Consider now the volume referring to the diffusion length at time $t$,
$V(t) = L(t)^{3} = (Dt)^{3/2}$. One can suppose that for lengths
larger than $L(t)$ the fluctuations of $z(\bi{r})$ are still not
affected by diffusion; therefore, one can write that
\begin{equation}
  \label{eq:zabs}
  \langle |z(\bi{r})| \rangle \approx
  \left(\bar{c}_{\mathrm{A}}^{0} / V(t)\right)^{1/2} =
  \left(\bar{c}_{\mathrm{A}}^{0}\right)^{1/2} D^{-3/4} t^{-3/4}.
\end{equation}
In some regions, $c_{\mathrm{A}}(\bi{r}) = |z(\bi{r})|$ and in others,
$c_{\mathrm{B}}(\bi{r}) = |z(\bi{r})|$; hence, on average
\begin{equation}
  \label{eq:solution2}
  \bar{c}_{\mathrm{A}}(t) = \bar{c}_{\mathrm{B}}(t) =
  \frac{1}{2} \langle |z| \rangle \approx
  \left(\bar{c}_{\mathrm{A}}^{0}\right)^{1/2} D^{-3/4} t^{-3/4},
\end{equation}
meaning a slower kinetics than in stage 1 \eref{eq:solution1}. (For a
more detailed derivation see \cite{ovchinnikov78}).

Note that \eref{eq:solution2} is only applicable within certain time
limits $[t_{1},t_{2}]$. For instance, for $k \to \infty$, $t_{1}$ is
equal to the time the diffusion length $L(t)$ needs to exceed the
average intermolecular distance. $t_{2}$ is determined by the time
needed by $L(t)$ to reach the system size $L_{\mathrm{s}}$,
independent of the value of $k$.

\subsection{Role of density fluctuations in the relaxation of random
  dislocation configurations}
\label{sec:application}

\subsubsection{Analogy between chemical and dislocation systems.}
\label{sec:densdisloc}

Now we consider the relaxation of initially random dislocation
configurations with the same number of positive and negative
dislocations ($\sum_{i} s_{i} = 0$) following the equations of motion
\eref{eq:eqm}. Although not immediately evident, this relaxation
process has strong phenomenological similarities to the kinetics of
the chemical reaction described in \sref{sec:kinetics}.

To elucidate the analogy, the relaxation process will be envisaged as
a gradual screening of the long range elastic stress
fields of individual dislocations through the formation of
dislocation--dislocation correlations \cite{zaiser01,groma06}. We
first envisage `neutral' arrangements where dislocations of both signs
are present in equal amounts. In a screened dislocation arrangement,
the excess of one sign over the other has been eliminated on scales
above a few dislocation spacings. Any dislocation arrangement where
excess dislocations are completely eliminated can be envisaged as an
assembly of dipoles where each dislocation has exactly one partner of
opposite sign within a distance of the order of $\rho^{-1/2}$, and
this picture will be used in the following argument. Hence, we
envisage the relaxation process as the gradual formation of a large
number of dislocation dipoles consisting of dislocations of opposite
signs, i.e., as a bimolecular reaction process analogous to the
chemical reaction \eref{eq:reaction}.

The principal difference between the two processes lies in the
dynamics of individual particles: for the dislocation system,
dislocation glide motion is driven by dislocation--dislocation
interactions which scale like $1/r$, whereas in case of the chemical
reaction we are dealing with diffusive Brownian motion of the
reactants. Despite this difference, we may again construct a two-stage
model for the dislocation relaxation process: i) in stage 1 adjacent
opposite sign dislocations form dipoles; ii) in stage 2 initial
fluctuations in the excess dislocation density gradually die out from
shorter towards longer length scales as excess dislocations which did
not find a dipole partner in stage 1 undergo long range glide
motion. The process terminates once the length scale on which
fluctuations have been eliminated reaches the system size.

\subsubsection{Relaxation dynamics.}
\label{sec:fundamentals}

To understand the time evolution of the system, we again consider the
evolution of the typical length scale $L(t)$ below which density
fluctuations have already died out. To this end we consider that i)
similar to the chemical case, areas of size $L(t)^{2}$ typically
contain mobile excess dislocations (which have still not `reacted'
into dipoles) with only one or the other
sign and that ii) dislocation dipoles give rise only to short range
stress fields with a $1/r^{2}$ decay. As the typical distance between
opposite sign dislocations which try to find each other is
proportional to $L'(t')$, the typical driving stress towards dipole
formation scales as $\tau' \sim 1/L'(t')$. (Recall that variables with
an apostrophe ($'$) are measured in natural units.) The dislocation
velocity is proportional to $\tau'$, and
therefore the characteristic time for eliminating excess dislocations
on scale $L'$ scales as $L'/\tau' \propto (L')^2$. Hence we find that
\begin{equation}
  \label{eq:Levo}
  \frac{\rmd L'}{\rmd t'} \sim \frac{1}{L'} \qquad \mathrm{or,
    equivalently} \qquad L' \sim \sqrt{2 t'}.
\end{equation}
Incidentally, this result is very similar to the evolution law of
$L(t)$ for Brownian motion, $L=\sqrt{Dt}$. By supposing that the
mobile dislocations inherit the initial concentration fluctuations we
find that at time $t'$, regions of size $L'(t')^{2} \sim 2 t'$ contain
about $L'$ excess dislocations of one or the other sign, and $L'^2$
dislocations in total (see the chemical model in
\sref{sec:diffusion}). Thus, the fraction of non-paired dislocations
is estimated to decrease in time as \begin{equation}
  \label{eq:excess}
  \frac{N_{\mathrm{excess}}}{N} \sim \frac{L'}{L'^{2}} \sim \frac{1}{\sqrt{2 t'}}.
\end{equation}

Following the chemical model in \sref{sec:diffusion}, it is
straightforward to predict the time interval $[t'_{1},t'_{2}]$ during
which the above argument is expected to hold. The start time $t'_{1}$
is characterised by $L$ reaching the dislocation--dislocation distance
$\rho^{-1/2}$ ($1$ in natural units) and the process is finished when
$L$ reaches the system size $L_{\mathrm{s}}$. With \eref{eq:Levo} this
leads to
\begin{equation}
  \label{eq:t12}
  t'_{1} \sim \frac{1}{2} \qquad
  t'_{2} \sim \frac{L_{s}'^{2}}{2} =
  \frac{\rho L_{\mathrm{s}}^{2}}{2} = \frac{N}{2} \qquad
  \frac{t'_{2}}{t'_{1}} = \frac{t_{2}}{t_{1}} \sim N
\end{equation}
in natural coordinates where $N$ denotes the total number of
dislocations in the system.

\subsubsection{Scaling relations for energy and velocity.}
\label{sec:pred}

The fraction of `non-paired' dislocations is not a convenient quantity
for comparing the scaling model with dislocation dynamics simulations,
as the definition of `dislocation pairs' in a multipolar dislocation
arrangement may be ambiguous. Instead, we consider the evolution of
the $n$th moment $\langle |v|^{n} \rangle$ of the dislocation velocity
and of the excess elastic energy, both of which can be determined from
the simulations in a straightforward manner. To obtain scaling
estimates for the relaxation of these quantities,  we assume that all
dislocation dipoles are at rest and only the excess dislocations
move. Furthermore, we assume that the motion of excess dislocations is
not hindered by the dipoles already formed (this can be rationalised
with the short range of the dipole stress field). The
velocity of excess dislocations scales as $|v'| \sim 1/L'$, leading to
\begin{equation}
  \label{eq:vmoment}
  \langle |v'|^{n} \rangle \sim
  \frac{N_{\mathrm{excess}}}{N} \times |v'|^{n} \sim
  \frac{1}{L'} \times \frac{1}{(L')^{n}} = (L')^{-(n+1)} \sim
  (2t')^{-(n+1)/2},
\end{equation}
where \eref{eq:Levo} and \eref{eq:excess} have been used.

The dynamics of dislocations is assumed to be overdamped. Hence, the
work that is done by the internal stresses in driving the system is
completely dissipated: the amount of dissipated energy exactly matches
the reduction in elastic energy. The energy dissipated per unit time
by a moving dislocation $i$ scales like $\tau_i v_i$ (the
Peach--Koehler force acting on the dislocation is proportional to the
stress), and consequently the time evolution of $E$  can be expressed
as
\begin{equation}
  \label{eq:DeltaEpre}
  \Delta E'(t') =
  - \int^{t'} \sum_{i=1}^{N}
  \tau'_{i}(t'') v'_{i}(t'') \rmd t''.
\end{equation}
Since the motion is overdamped, the dislocation velocity is
proportional to the acting stress. In natural units and for $t' \in
[t_{1}', t_{2}']$, the ensemble averaged elastic energy thus evolves
like
\begin{equation}
  \label{eq:DeltaE}
  \eqalign{
    \left\langle \Delta E'
    \right\rangle \propto
    &- \int^{t'} \sum_{i=1}^{N}
    v_{i}'^2(t'') \rmd t'' \approx - N
    \int^{t'} \langle v'^{2} \rangle \rmd t''
    \approx \cr
    &- N \int^{t'} (2 t'')^{-3/2} \rmd t'' =
    N/\sqrt{2t'} ,
  }
\end{equation}
where \eref{eq:vmoment} was used to estimate the time evolution of the
second velocity moment.

\subsubsection{Relaxation of arrangements of dislocations of the same sign.}
\label{sec:single2}

The scaling argument in the previous section is based on the formation
of dipoles consisting of edge dislocations of opposite signs. At first
glance, such an argument seems to be completely inapplicable to
systems where only dislocations of the same sign are present. Dipole
formation is clearly impossible in such systems. Instead, a most
conspicuous feature in the relaxation of single-sign edge dislocation
systems is the formation of walls containing many dislocations of the
same sign that are  aligned in the direction perpendicular to the slip
plane \cite{thomson06}. Accordingly, theoretical arguments have
focused on parameters characterising the `condensation' of
dislocations into walls \cite{thomson06}. However, even though
dislocation wall formation is a most conspicuous feature, wall
formation alone can \emph{not} produce a screened dislocation
arrangement. The authors of \cite{thomson06} evaluate the driving
force for wall formation by assuming periodic spacing of dislocations
along a wall and using classical results found e.g.\ in
\cite{hirthlothe82}. If we follow this line of reasoning and note that
the energy of a dislocation in a wall decreases with decreasing
dislocation spacing, the minimum-energy structure for the system at
hand would be a single system-spanning wall. However, for an initially
random dislocation system the $y$ positions
are independent random variables and it is not easy to see how a
periodic arrangement could form in the absence of dislocation
climb. We calculate the energy of a random wall in the Appendix and
show that forming such a wall does not produce any energy reduction
with respect to the initial random 2D arrangement.

How then can an arrangement of dislocations of the same sign be
screened? The answer was provided by Wilkens \cite{wilkens69} who
demonstrated that a screened dislocation arrangement can be
constructed by eliminating dislocation density fluctuations above a
certain scale. To this end, he proposed a construction where the
crystal cross section is tiled into a grid of cells of size $l$, and
the same number of dislocations is randomly distributed within each
cell. This construction, which eliminates all density fluctuations on
scales above the cell size, leads to an arrangement where the
screening radius coincides with the cell size, ${\cal E} \propto \rho \ln
(l/b)$.

Taking the Wilkens construction as a well-screened reference state
offers a surprising outlook on the relaxation of initially random
systems of same-sign dislocations. With respect to this reference
state, the initial random arrangement contains density fluctuations on
all scales which may be either positive ($\delta \rho = \rho(\bi{r}) -
\langle \rho \rangle > 0$, positive excess) or negative ($\delta \rho
<0$,  negative excess). To achieve screening, dislocations must
migrate from regions of positive to regions of negative excess, and
this process is governed by the long-range stress fields associated
with the presence of (positive or negative) excess dislocations. In
other words, the kinetics of the process follows from
exactly the same scaling argument as used in the previous section: We
are dealing with the stress-driven elimination of excess dislocation
densities, with the only difference that the excess is now not of
positive over negative dislocations, but of the local dislocation
density over the average one.

If the above argument is correct, the relaxation kinetics of same-sign
dislocation systems should be characterised by a slow power-law stage
which has the same characteristics as the relaxation of
neutral dislocation systems as discussed in the previous
sections. We demonstrate in the next section that this is indeed the
case.

\section{Comparison with simulation results}
\label{sec:comparison}

\subsection{Single slip relaxation of dislocation arrangements with
  zero net Burgers vector}
\label{sec:single1}

\subsubsection{Evolution of the elastic energy.}
\label{sec:energy}

\begin{figure}
  \centering
  \includegraphics{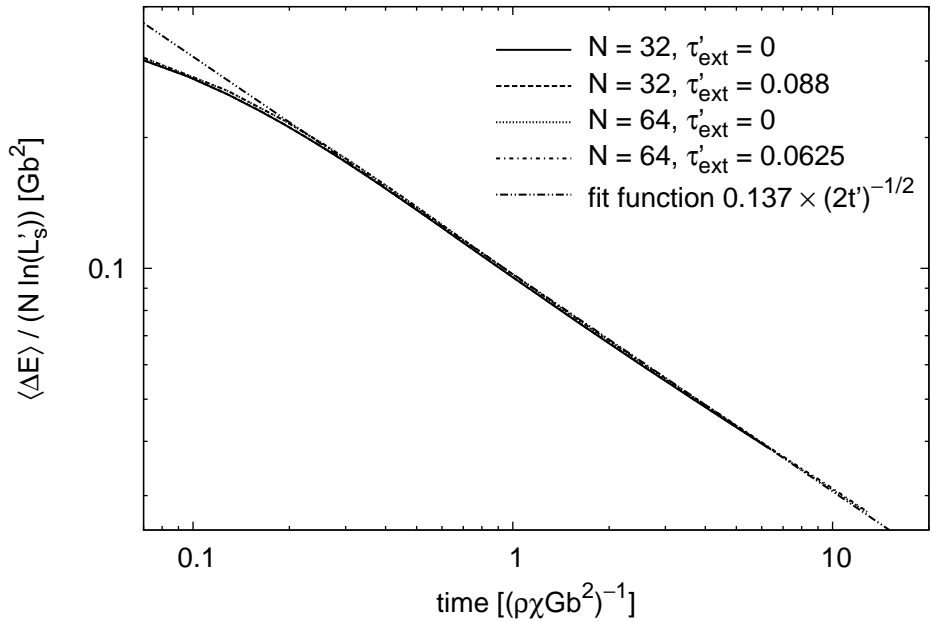}
  \caption{Evolution of the elastic energy per dislocation for
    different system sizes $N$ and external stress values
    $\tau'_{\mathrm{ext}}$.}
  \label{fig:DeltaE}
\end{figure}

The numerically determined evolution of the elastic energy is
displayed in \fref{fig:DeltaE}. The zero value of the energy was
chosen to correspond to the final relaxed state of the system. As seen
on the figure, the evolution of the elastic energy $\Delta E'/N$ per
dislocation can be fitted satisfactorily with the prediction in
\eref{eq:DeltaE}, $\Delta E/N = A / \sqrt{2t'}$. Power-law
relaxation occurs from times $t'_{1} \approx 0.3$ onwards, in good
agreement with the model prediction $t'_{1} \sim 1/2$ in
\eref{eq:t12}. The final equilibrium value of the elastic energy
is not a priori known but was fitted to the data such as to achieve a
maximum extension of the linear scaling regime. Unfortunately this
precludes determination of the second critical time $t'_{2}$ from
these results. The presence or absence of an external stress below the
flow stress of the relaxed dislocation system seems to have negligible
influence on the evolution of the elastic energy.

Note that it is possible to collapse the curves for different system
sizes $L'_{\mathrm{s}}$ by normalising the graphs with
$\ln(L'_{\mathrm{s}})$, as done in \fref{fig:DeltaE}. This observation
is consistent with the fact that the elastic
energy of the initial random dislocation system is of the order of
$E_{0} \sim N G b^{2} \ln(L_{\mathrm{s}}/b)$ while the energy of the
final relaxed state is of the order of $E_{\infty} \sim N G b^{2}
\ln((\sqrt{\rho} b)^{-1})$ \cite{wilkens67,wilkens69}. For estimating
the value of $E_{\infty}$ we used that the range of dislocation pair
correlations in mechanical equilibrium is of the order of the mean
dislocation--dislocation distance $\rho^{-1/2}$
\cite{zaiser01}. Therefore, the elastic energy difference per
dislocation between the initial and final states is of the order of
\begin{equation}
  \label{eq:DeltaEWilkens}
  \Delta E / N =
  (E_{0} - E_{\infty}) / N \sim
  G b^{2} \ln(L_{\mathrm{s}}\sqrt{\rho}) =
  G b^{2} \ln(L_{\mathrm{s}}').
\end{equation}
Despite this relation connecting the initial and final states of the
system, the numerical finding that the
$\Delta E(t)/N$ curves for different system sizes can be collapsed on
their entire course by normalising them with
$\ln(L'_{\mathrm{s}})$ is not trivial, as the agreement extends also
to the relaxation kinetics and characteristic crossover time $t'_1$.

\subsubsection{Evolution of the mean square velocity.}
\label{sec:v2}

\begin{figure}
  \centering
  \includegraphics{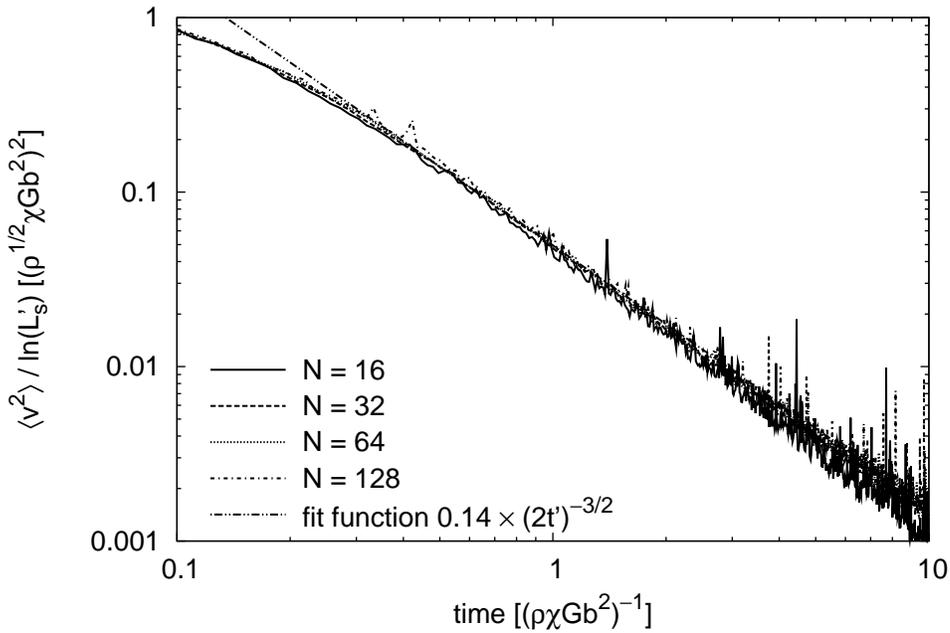}
  \caption{Evolution of the mean square velocity
    $\langle v'^{2}\rangle$ for different system sizes $N$ at zero
    external stress.}
  \label{fig:v2}
\end{figure}

The numerically calculated evolution of the mean square velocity
$\langle v'^{2} \rangle$ is displayed in \fref{fig:v2} for zero
applied stress and different system sizes. Due to the connection
\eref{eq:DeltaE} between the mean square velocity and the
elastic energy of the system, it is not surprising that similar
statements apply here as for the evolution of the elastic
energy. As seen in the figure, the model prediction $\langle v'^{2}
\rangle \sim (2t')^{-3/2}$ in \eref{eq:vmoment} fits
the data well from $t'_{1} \approx 0.3$. Due to the fact that $\langle
v'^{2} \rangle$ is proportional to the time derivative of the elastic
energy, its graphs are much noisier than those obtained for the
energy, preventing again the detection of the supposed upper critical
time $t'_{2}$. As for the energy, size effects can be scaled out with
a normalisation factor $\ln(L'_{\mathrm{s}})^{-1}$ which is a direct
consequence of \eref{eq:DeltaE} and \eref{eq:DeltaEWilkens}. A final
analogy to the evolution of the elastic energy is that external
stresses have only negligible influence on the evolution of the mean 
square velocity. For this reason, simulations with non-zero external 
stresses were omitted from \fref{fig:v2}.

\subsubsection{Evolution of the mean absolute velocity.}
\label{sec:vabs}

\begin{figure}
  \centering
  \includegraphics{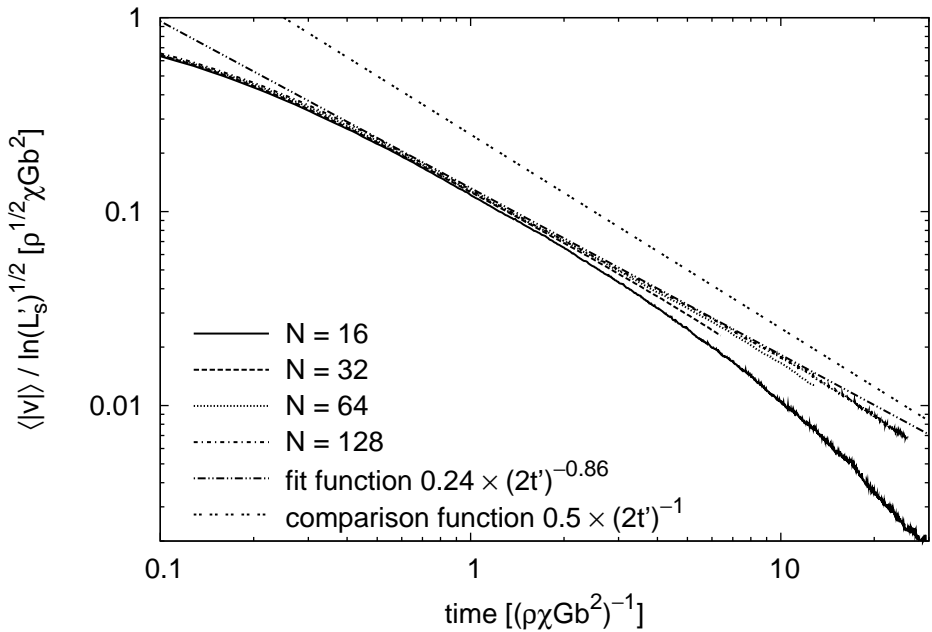}
  \caption{Evolution of the mean absolute velocity
    $\langle{}|v'|\rangle$ for different system sizes $N$ at zero
    external stress.}
  \label{fig:vabs}
\end{figure}

In \fref{fig:vabs} the evolution of the mean absolute velocity
$\langle |v'| \rangle$ can be seen for different system sizes. Again,
a power law time dependence can be observed from $t'_{1} \approx 0.3$
although an exponent $-0.86$ gives a better fit than the theoretically
predicted $-1$ expected according to equation \eref{eq:vmoment}. One
may argue that slowly moving dislocation dipoles play a bigger role in
this case, as their small velocities contribute more strongly to
$\langle |v'|\rangle$ than to $\langle v'^2 \rangle$. Therefore, the
gradually increasing number of slowly moving dislocations might be
responsible for the reduced relaxation exponent. What makes this
figure very interesting is the possibility to estimate values of
$t'_{2}$. It was found that $t'_{2} \approx 0.2 N$ gives a good
approximation, in line with the model prediction $t'_{2} \sim N/2$ in
\eref{eq:t12}. It was also observed that normalisation with
$(\ln(L'_{\mathrm{s}}))^{-1/2}$ collapses the graphs referring to
different system sizes $L'_{\mathrm{s}}$ in the region of small $t' \ll
t'_2$. This is consistent with the relations for the elastic energy
and the mean square velocity. Finally, as in case of the energy and the
mean square velocity, the evolution of the mean absolute velocity
is not changed by the presence of external stresses below the macroscopic 
flow stress. 

\subsubsection{Evolution of the plastic strain rate.}
\label{sec:strainrate}

Another numerically measurable quantity is the plastic strain rate, defined as
\begin{equation}
  \label{eq:gammdotdef}
  \dot{\gamma} = \sum_{i} \frac{b_{i} \dot{x}_{i}}{L_{\mathrm{s}}^{2}}
  = b L_{\mathrm{s}}^{-2} \sum_{i} s_{i} \dot{x}_{i} \qquad
  \dot{\gamma}' = N^{-1} \sum_{i} s_{i} \dot{x}_{i}'.
\end{equation}
In the following the evolution of $\dot{\gamma}'$ is studied for
applied shear stresses $\tau_{\mathrm{ext}}'$ below the macroscopic
flow stress for the present dislocation geometry. From the data of
Miguel and co-workers \cite{miguel02,miguel07}, this is estimated to
be $\sim 0.1$ in natural units. As it was demonstrated in
\fref{fig:DeltaE}, external stresses in this range do not appreciably
change the evolution of the elastic energy. We study the relaxation of
the strain rate mainly in order to assess, by comparing with the work
of Miguel et al., the relevance of different initial conditions on
the creep behaviour of dislocation systems.

\begin{figure}
  \centering
  \includegraphics{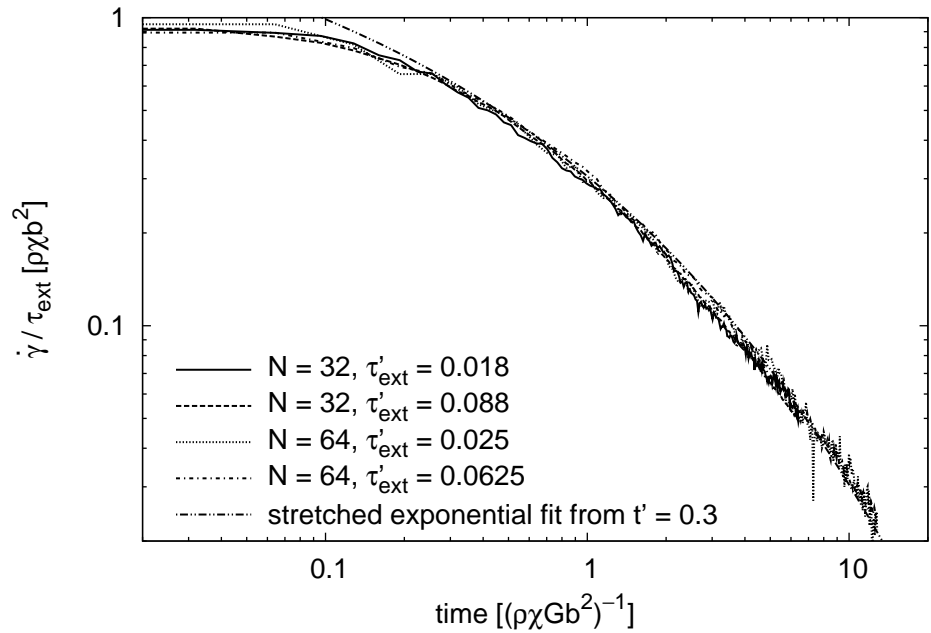}
  \caption{Evolution of the plastic strain rate $\dot{\gamma}'$
    normalised with the external stress $\tau'_{\mathrm{ext}}$ for
    different system sizes $N$ and external stress values
    $\tau'_{\mathrm{ext}}$.}
  \label{fig:gammadot}
\end{figure}

\Fref{fig:gammadot} shows the numerically determined evolution of the
plastic strain rate $\dot{\gamma}'$ for different system sizes and
external stress values. As can be seen, the plastic strain rate scales
roughly in proportion with the external
stress. The relaxation does not follow any discernible power law but
is roughly exponential. This is in marked contrast with the findings
of Miguel et al.\ \cite{miguel02,miguel07} who for well relaxed
initial configurations demonstrate an Andrade-type power-law decay,
$\dot{\gamma} \propto t^{-2/3}$. The discrepancy points to the crucial
importance of initial conditions for relaxation processes in
dislocation systems -- a factor which is also borne out by the history
dependence of creep relaxation processes that was demonstrated by
Miguel et al.\ \cite{miguel07}.

\subsubsection{Role of boundary conditions (simulation box
  orientation).}
\label{sec:phi}

\begin{figure}
  \centering
  \includegraphics[width=0.8\textwidth]{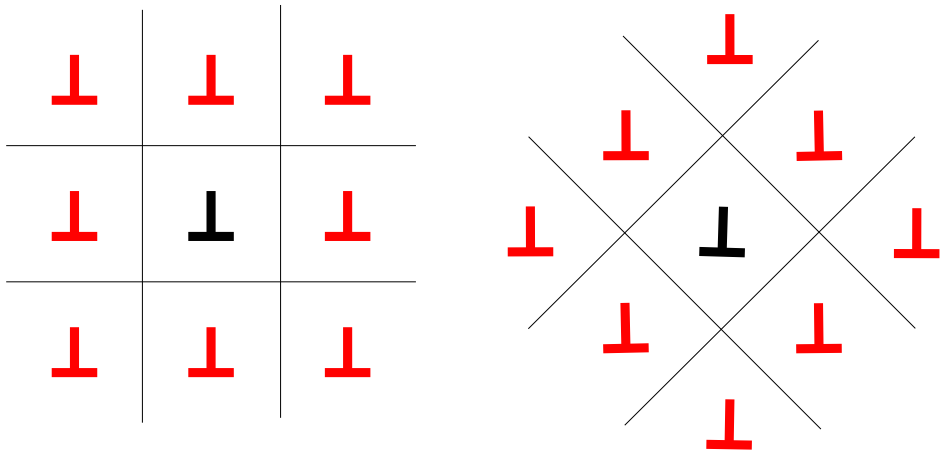}
  \caption{Arrangement of the first 8 periodic images of a given dislocation
  	for $\varphi = 0^{\circ}$ (left) and $\varphi = 45^{\circ}$
        (right). The latter configuration has a higher elastic energy
        as the nearest neighbours of each dislocation are in an
        energetically unfavourable configuration.}
  \label{fig:images}
\end{figure}

In this section we investigate the influence of different ways of 
implementing the periodic boundary conditions by tilting the angle
between the edges of the simulation box and the trace of the slip 
planes. Simulations with different tilt angles $\varphi$ are physically 
equivalent except for the spatial arrangement of the periodic images of each
dislocation (see \fref{fig:images}). This arrangement affects the 
dislocation--dislocation interactions on scales comparable to the 
simulation box size.  Also, the interaction energy of each dislocation 
with its periodic images affects the initial elastic energy of the
system, which is smallest for $\varphi=0$ and has a maximum for 
$\varphi = 45^{\circ}$.

\begin{figure}
  \centering
  \includegraphics{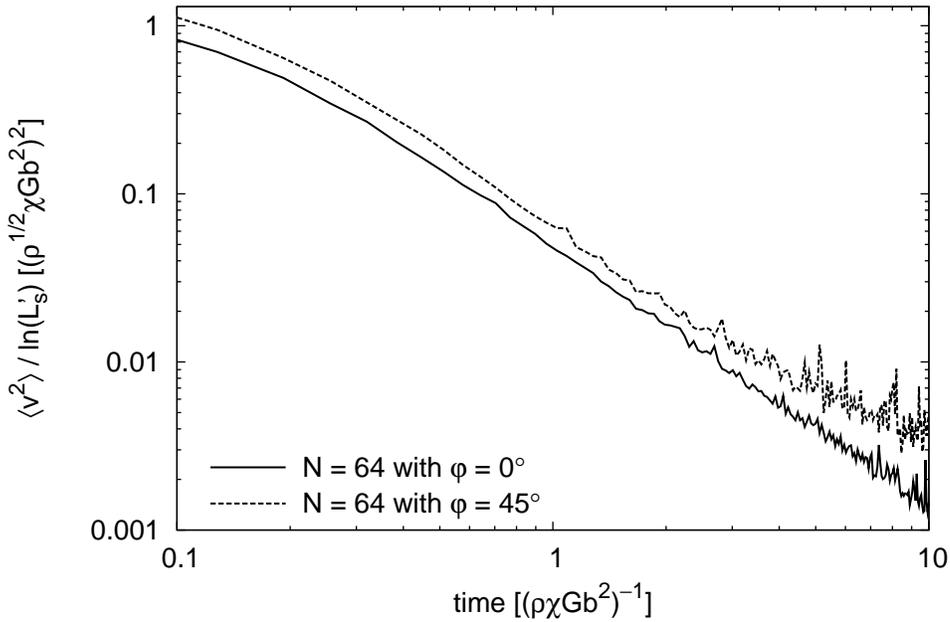}
  \caption{Evolution of the mean square velocity
    $\langle v'^{2}\rangle$ in systems with a single slip geometry
    with the simulation box oriented at different angles $\varphi$ to
    the slip planes.}
  \label{fig:phi}
\end{figure}

The influence of simulation box orientation on the relaxation process 
is illustrated in \fref{fig:phi}. The absolute
values of the squared velocity (or equivalently the energy dissipation
rate) are higher for $\varphi = 45^{\circ}$ than for $\varphi =
0$. However, both curves differ only by a constant factor (the ratio
of the initial excess energies), while the dynamics of the relaxation
processes is otherwise identical.

\subsection{Relaxation of arrangements of dislocations of the same sign}

For systems of dislocations of the same sign we have numerically
evaluated the evolution of the mean square velocity $\langle v'^{2}
\rangle$ (or, equivalently, of the energy dissipation rate) and of the
mean absolute velocity $\langle |v'| \rangle$. All calculations were
performed at zero external stress since any applied stress would
induce a sustained drift motion of the dislocation arrangement.

\begin{figure}
  \centering
  \includegraphics{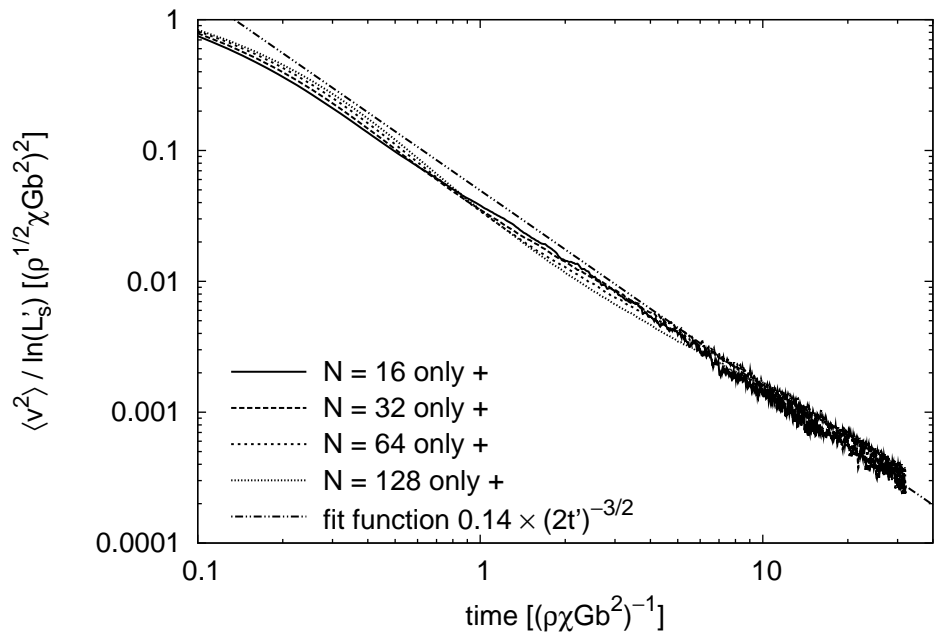}
  \caption{Evolution of the mean square velocity
    $\langle v'^{2}\rangle$ for different system sizes $N$ for
    dislocations of the same sign.}
  \label{fig:v2same}
\end{figure}

\begin{figure}
  \centering
  \includegraphics{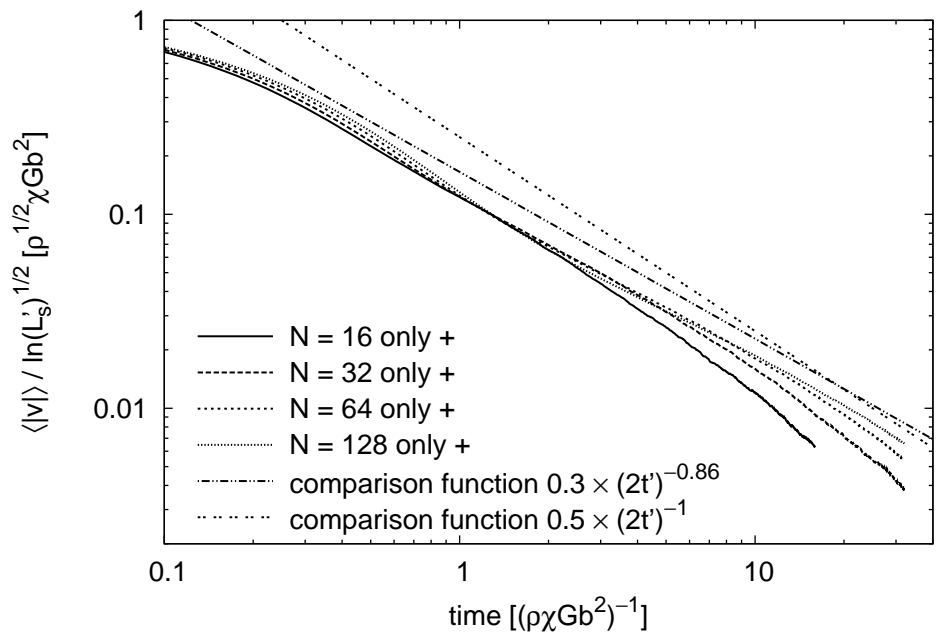}
  \caption{Evolution of the mean absolute velocity
    $\langle{}|v'|\rangle$ for different system sizes $N$ for
    dislocations of the same sign.}
  \label{fig:vabssame}
\end{figure}

Results are shown in  \fref{fig:v2same} and \fref{fig:vabssame}
together with the fit functions obtained for neutral dislocation
arrangements (see \fref{fig:v2} and \fref{fig:vabs}). It is evident
that the relaxation of same-sign dislocation systems follows
the same scaling laws that have been observed for systems containing
equal numbers of dislocations of both signs. This provides strong
support for our basic conjecture that relaxation is governed by the
stress-driven elimination of excess dislocations in a process that
progresses from small to large scales. The processes occurring on short
scales, on the other hand, are evidently different for the two systems
(dipole formation vs.\ formation of walls). This is reflected
by the fact that the relaxation process in the single-sign dislocation systems
shows an initial size dependence which is not present in 
neutral dislocation systems (see \fref{fig:v2} and \fref{fig:vabs} for comparison).

\subsection{Dislocation relaxation on multiple slip systems}
\label{sec:multi}

\begin{figure}
  \centering
  \includegraphics{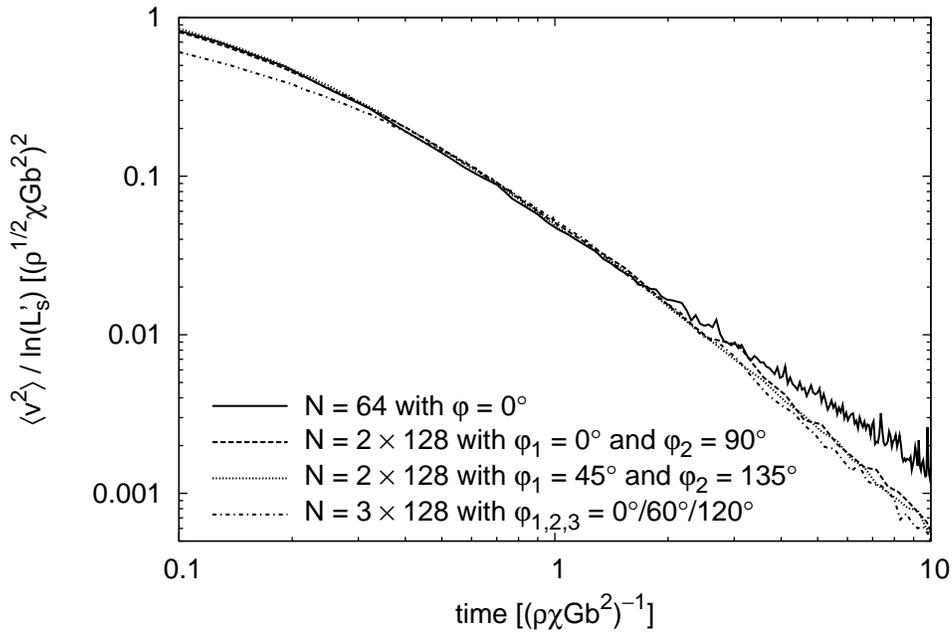}
  \caption{Comparison of the evolution of the mean square velocity
    $\langle v'^{2}\rangle$ in dislocation systems with single and
    multiple slip geometries. In the figure, $\rho$ means the total
    density of dislocations on all slip systems. The last two curves
    were shifted downwards to help comparison.}
%    and
%    $\alpha(0^{\circ}) = \alpha(0^{\circ}/90^{\circ}) = 1$,
%    $\alpha(45^{\circ}/135^{\circ}) = \alpha(45^{\circ}) = 3$ and
%    $\alpha(0^{\circ}/60^{\circ}/120^{\circ}) = 200$ were chosen to
%    define the effective system size
%    $L_{\mathrm{s, eff}}' = \alpha(\varphi) \sqrt{N}$.}
  \label{fig:multi}
\end{figure}

\Fref{fig:multi} compares the relaxation of dislocation systems in
single, double and triple slip. As seen in the figure, at long times
the relaxation in multiple slip geometries accelerates in comparison
with the relaxation in single slip.
This is consistent with the idea that in multiple slip geometries
relaxation processes proceed through the formation of dislocation
locks and the annihilation of mobile dislocations at these locks. The
long range stress fields associated with dislocation locks and the
removal of dislocations accelerate the relaxation as seen in
the figure. One of the main assumptions behind our scaling model,
namely that the motion of mobile dislocations is governed mainly by
their mutual interaction, no longer holds in multiple slip
geometries. Therefore, the scaling model can not be applied to these
situations. Indeed, as seen in \fref{fig:multi}, no power law
relaxation regime can be detected in the multiple slip simulations.

\section{Discussion and conclusions}
\label{sec:conclusions}

The present paper discusses the relaxation of initially random
arrangements of straight, parallel edge dislocations. Following a
phenomenological analogy with the kinetics of bimolecular reactions
\cite{ovchinnikov78} the relaxation process can be divided into three
consecutive stages. Stage 1 is characterised by rapid rearrangements
of neighbouring dislocations, leading to the formation of dipoles,
multipoles, and/or short wall segments. Stage 2 hosts the gradual
extinction of initial fluctuations in the Burgers vector density on
ever increasing length scales through the long range transport of
excess dislocations. This stage gives rise to characteristic power-law
relaxation dynamics. The relaxation process terminates in stage 3 when
the characteristic fluctuation length reaches the system size.

Our considerations focus on the power-law relaxation dynamics in stage
2. In section 3, we formulated a scaling theory for this process by
making the analogy with a bimolecular reaction. To this end, we
considered a highly simplified picture where pair  interactions
between positive and negative excess dislocations lead to long-range
dislocation transport resulting in the formation of dislocation
dipoles. This led to predictions for the evolution of the elastic
energy and the first two moments of the dislocation velocity. These
predictions were then compared to ensemble averages of discrete
dislocation dynamics simulations, and convincing agreement was found
for single slip geometries. For multiple slip geometries, however, the
persistent long range stress fields of dislocation locks accelerate
the relaxation process, to which the scaling model can no longer be
applied.

The actual dislocation processes in many-dislocation systems are much
more complex than the simplified picture underlying our scaling
arguments. Instead of long-range transport of excess dislocations and
dipole formation, we see complex rearrangements resulting in
dislocation dipoles, multipoles, and walls. In addition, it is well
known that dislocations have a propensity to form large-scale
heterogeneous patterns consisting of dislocation-rich and dislocation
depleted regions. In the following we briefly discuss how these
complex static and dynamic features fit into the idealised picture we
used in the previous sections.

\subsection{Dipoles, multipoles and walls}

We have developed our scaling argument for dipole formation which can
be considered a bimolecular reaction between positive and negative
dislocations. However, actual dislocation arrangements are much more
complex. It is therefore important to emphasise that the core of the
argument is the elimination of large-scale fluctuations in the excess
Burgers vector density, and \emph{not} the resulting arrangement of
nearby dislocations. For the long-time asymptotics of the
relaxation process, which is governed by the elimination of fluctuations on
larger and larger scales, the small-scale arrangement of dislocations is 
virtually irrelevant -- at least as long as the local features (dipoles, 
multipoles, short walls and combinations of all these) do not give rise 
to long-range stresses. According to the present argument, a well-screened 
dislocation arrangement is one in which
Burgers vector fluctuations have been eliminated on all scales. If we
are dealing with a neutral dislocation system (equal  numbers of
positive and negative dislocations) this means that the net Burgers
vector is zero in each small volume, i.e., for each dislocation we can
find exactly one partner of opposite sign nearby. This motivates the
dipolar picture even if the actual dislocation arrangements may be
more complex.

An alternative mechanism for creating well-screened dislocation
arrangements is the formation of system-spanning walls of dislocations
of the same sign. Periodic arrangement of edge dislocations of one
sign into a wall perpendicular to the glide plane removes the
logarithmic divergence of the dislocation energy and introduces a
screening length that is proportional to the dislocation spacing along
the wall \cite{hirthlothe82}. Walls are conspicuous both in
simulations and in many  experimentally observed dislocation
microstructures. At first glance, wall formation mechanism seems
to be completely at odds with the mechanism discussed in the present paper:
Formation of walls of same-sign dislocations increases, rather than
reduces, the Burgers vector density fluctuations. However, a closer
investigation reveals that wall formation by itself is not a screening
mechanism at all. Forming a wall of randomly spaced dislocations does
not reduce the energy in comparison with a random 2D dislocation
arrangement (see Appendix). Instead, the screening effect is
contingent on the equal spacing of dislocations, i.e.\ on suppressing 
fluctuations of the Burgers vector density along the wall direction. 
If we start from a random dislocation arrangement this is not easy to 
obtain: Either the dislocations must
have climb degrees of freedom (which we do not consider in the present
study), or dislocation motions that lead to the formation of multiple
walls must be correlated over large distances such as to ensure that
each wall collects only those dislocations that fit into an evenly
spaced pattern. In the latter case, we are again dealing with the
suppression of Burgers vector density fluctuations on all scales above
the wall spacing, and the long time asymptotics of this process is
expected to obey our scaling theory.
This is confirmed by the simulations, which however also demonstrate
that the short-time behaviour is different for neutral dislocation
systems where the local arrangement of dislocations is characterised
by dipolar and multipolar patterns, and for single-sign dislocation
systems where the local arrangement of dislocations is characterised
by walls (compare \fref{fig:v2} and \fref{fig:v2same}).

\subsection{Long-range transport of excess dislocations}

\begin{figure}
  \centering
  \includegraphics[width=0.8\textwidth]{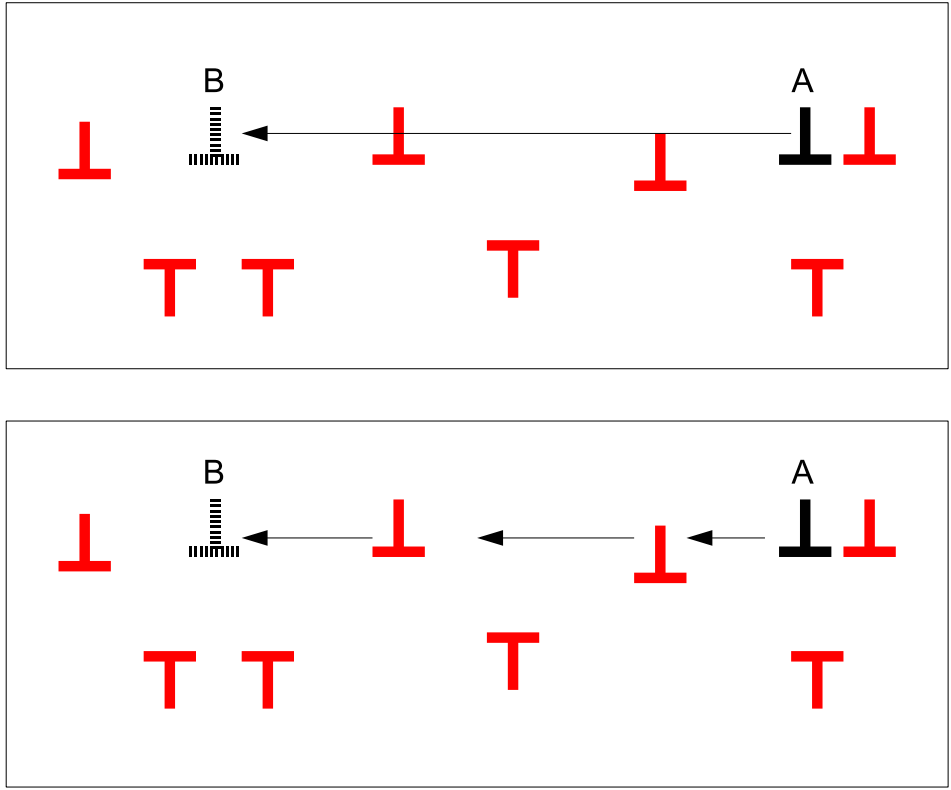}
  \caption{Schematic illustration of excess dislocation transport in a
    multipolar dislocation arrangement: a positive excess
    dislocation at A `recombines' with a negative excess dislocation
    at B  (top) by direct motion from A to B or (bottom) by
    collective rearrangement involving multiple dislocations.}
  \label{fig:multipole}
\end{figure}

Our scaling argument considers the stress-driven long-range transport
of excess dislocations. The picture underlying the argument is
schematically shown in \fref{fig:multipole} (top): A positive excess
dislocation at A is attracted by a negative
excess dislocation at B and the two recombine by long-range motion
which is not affected by the stress field of the dislocations in
between A and B. The figure also indicates that this idealisation may
not be feasible when we are dealing with multipolar arrangements
rather than isolated narrow dipoles: In that case the mutual
interaction of the excess dislocation may be much weaker than their
interaction with other dislocations `on the way'. As a consequence,
recombination is much more likely to occur by a collective
rearrangement as shown in \fref{fig:multipole} (bottom).

How does this affect our scaling argument? The total driving force for
the process is the same in both cases. However, a collective
rearrangement on scale $L$ is likely to involve $N = L\sqrt{\rho}$
dislocations [$N=3$ in \fref{fig:multipole} (bottom)]. Hence the
driving force per dislocation is reduced by a factor of the order of
$N$. However, the same is true for the characteristic distance that
has to be covered by each dislocation ($L$ in the case of direct
transport, $1/\sqrt{\rho}$ in the case of collective
rearrangement). As we assume that the dislocation velocity is
proportional to the driving force, it follows that the characteristic
time scale for eliminating the excess dislocation is the same in both
cases, and our scaling argument remains valid.

\subsection{Large-scale dislocation patterning}

It is a well known phenomenon that dislocation microstructures forming
during plastic deformation form heterogeneous patterns consisting of
regions of high and low dislocation density, with characteristic lengths 
that are large in comparison with the dislocation spacing. On a conceptual 
level, possible mechanisms underlying this patterning were discussed by Nabarro
\cite{nabarro00} who pointed out that it may be energetically
favourable to `segregate' the dislocation microstructure into areas of
high and low density: If we are dealing with a well-screened
dislocation arrangement, the energy density scales like ${\cal E}
\propto \rho \ln(1/(b\sqrt{\rho}))$. In this case it can be easily
shown that it is energetically favourable to increase $\rho$ in some
regions and decrease it proportionally in others. Such `energetically
driven' dislocation patterning could be a reason for the formation of 
dislocation-dense and dislocation-depleted regions that is observed 
in many experiments. While this mechanism is not covered by the present 
model, it is not at variance with our considerations: In a neutral dislocation
arrangement, the formation of dislocation-dense and
dislocation-depleted regions might occur without disturbing the
Burgers vector balance. We note, however, that in our simulations
large-scale dislocation patterning is not observed -- either because it
does indeed not occur in single slip, or because the dislocation
numbers in our simulations might be too small.

\subsection{Conclusion}

In conclusion we discuss the relevance of the processes discussed in
the present paper for real-world systems. The relaxation of a random
dislocation system has no direct counterpart in real deformation
experiments, since it is impossible to `prepare' such a random
system in the first place. Our analysis of the screening of same-sign
dislocation systems is, however, of general importance for
understanding real dislocation patterns since it demonstrates that
wall formation, though a conspicuous feature, can by itself not
account for screening. This observation points to the importance of
investigating long-range correlations between dislocation positions both 
within the walls and across different walls, and offers ample scope 
for future investigations.

The investigated processes are of significant importance for discrete
dislocation dynamics simulations of plasticity as the slow nature of
the relaxation makes it difficult to obtain well-defined and
energetically stable initial configurations. Our comparison of
strain-rate relaxation experiments with the results of Miguel et
al.\ demonstrates that the collective behaviour of dislocation systems
may depend significantly on initial conditions. The analysis of this
dependence is still in its infancy, yet understanding it is
indispensable for carrying out dislocation plasticity simulations in a
controlled and well-defined manner.

\ack
Financial support of the Hungarian Scientific Research Fund (OTKA)
under Contract No.\ K 67778, of the European Community's Human
Potential Programme under Contract Nos.\ MRTN-CT-2003-504634
[SizeDepEn] and NMP3-CT-2006-017105 [DIGIMAT] and of
NEST Pathfinder programme TRIGS under contract
NEST-2005-PATH-COM-043386 are gratefully acknowledged.

\appendix
\section*{Appendix: Calculation of the elastic energy of a random
  dislocation wall}
\setcounter{section}{1}

We consider a wall of infinite height running along the plane $x=0$ in
an isotropic material. Edge dislocations of Burgers vector $\bi{b} = b
\bi{e}_{\mathrm{x}}$ are distributed randomly along the wall with average linear
density $1/h$. The geometry corresponds to a plane-strain situation,
hence the elastic energy density can be written as
\begin{equation}
{\cal E} = \frac{1}{4 \mu} \left[ (\sigma_{xx} +
  \sigma_{yy})^2(1-\nu)+2(\sigma_{xy}^2-\sigma_{xx}\sigma_{yy})
\right].
\label{edens}
\end{equation}
The total energy of the system is obtained by integrating
\eref{edens} over the system volume,
\begin{equation}
E = \int_V {\cal E} {\rm d}^2 r = \frac{1-\nu}{4 \mu} \int_V
(\sigma_{xx} + \sigma_{yy})^2\, {\rm d}^2 r,
\label{etot}
\end{equation}
where we have used that, for an infinite system, the second and third
terms on the right-hand side do not contribute to the total
energy. This can be shown as follows: For plane-strain deformation,
the stresses can be written as derivatives of the Airy stress function
$\chi$, $\sigma_{xx} = \partial_y^2 \chi$, $\sigma_{yy} = \partial_x^2
\chi$ and $\sigma_{xy} = -\partial_x\partial_y \chi$. Hence,
\begin{equation}
\int_V [\sigma_{xy}^2-\sigma_{xx}\sigma_{yy}] {\rm d}^2 r =  \int_V
[\partial_x\partial_y\chi \partial_x \partial_y \chi
- \partial_x^2\chi \partial_y^2 \chi ] {\rm d}^2 r.
\label{airy}
\end{equation}
Partially integrating the second term in the integral on the
right-hand side with respect to $x$ and $y$ shows that this integral
contributes only surface terms to the total energy. These terms are
negligible in the infinite-system limit.

The ensemble-averaged stress at any point is given by summing over the
stress fields of the individual dislocations in the wall and averaging
over the different realizations of the random dislocation positions:
\begin{equation}
\langle \sigma_{ij}(\bi{r}) \rangle = \sum_n \langle
\sigma_{ij}^{(n)}(\bi{r}) \rangle,
\end{equation}
where $\sigma_{ij}^{(n)}$ is the $ij$ component of the stress
created at $\bi{r}$ by the $n$th dislocation. The elastic energy of
the system depends on the averages of products  $\langle
\sigma_{ij}(\bi{r})\sigma_{kl}(\bi{r})\rangle$ where ${\it (ij,kl)}
\in [{\it (xx,xx); (xx,yy); (yy,yy)}]$. In evaluating these averages
we use that the $y$ coordinates of the individual dislocations are independent
random variables:
\begin{eqnarray}
\langle \sigma_{ij}(\bi{r})\sigma_{kl}(\bi{r})\rangle &=& \langle
\sigma_{ij}(\bi{r}) \rangle\langle \sigma_{kl}(\bi{r})
\rangle\nonumber\\ &+& \sum_n \left[\langle \sigma_{ij}^{(n)}(\bi{r})
  \sigma_{kl}^{(n)}(\bi{r}) \rangle
- \langle \sigma_{ij}^{(n)}(\bi{r}) \rangle \langle
\sigma_{kl}^{(n)}(\bi{r}) \rangle \right].
\label{indep}
\end{eqnarray}
We now make the following observations:
\begin{itemize}
\item
The average stresses $\langle \sigma_{ij}(\bi{r}) \rangle$ and their
products $\langle \sigma_{ij}(\bi{r})\sigma_{kl}(\bi{r})\rangle$
depend on the $x$ coordinate only.
\item
The average single-dislocation stresses $\langle \sigma_{xx}^{(n)}
\rangle$ and $\langle \sigma_{yy}^{(n)} \rangle$  become zero in the
limit $a \to \infty$, since these stresses are antisymmetric functions
of the $y$ coordinate. The same is true for the average total stresses
$\langle \sigma_{xx}\rangle$ and $\langle \sigma_{yy} \rangle$.
\end{itemize}
With these observations and using \eref{indep} we can write the
system energy as
\begin{eqnarray}
E = \int_V {\cal E} {\rm d}^2 r &=& \sum_n \frac{1-\nu}{4 \mu} \int_V
\langle (\sigma_{xx}^{(n)} + \sigma_{yy}^{(n)})^2 \rangle {\rm d}^2 r
\nonumber\\
&=& N  Gb^2\pi(1-\nu)
\ln\left[\frac{L}{2b}\right],
\label{efinal}
\end{eqnarray}
where the second step follows by interchanging the averaging and the
integration. $L$ is the system size (tending to infinity) which, in the absence of
any other screening mechanism, delimits the divergence of the dislocation self-energy. 
For periodic boundary conditions, as used in our simulations, $L$ must be understood 
as the size of the periodic simulation box which in this case defines the screening 
length for an otherwise uncorrelated dislocation arrangement. $N=L/h$ is the total 
number of dislocations in the system. 

It follows from \eref{efinal} that the energy per dislocation is equal 
to the energy of a single unscreened dislocation and, hence, equals the 
energy in a completely random 2D arrangement. In other words, the 
arrangement of dislocations of the same sign in a random wall does 
(with the possible exception of surface terms that are negligible in 
the infinite system limit) not produce any reduction of the total 
energy. Accordingly, the thermodynamic driving force towards forming 
such a wall is zero.


\Bibliography{99}
\bibitem{zaiser01b} Zaiser M, {\it Statistical modelling of dislocation
    systems}, 2001 {\it Mat. Sci. Eng.} A {\bf 309--310} 304--15
\bibitem{groma03} Groma I, Csikor F F and Zaiser M, {\it Spatial
    correlations and higher-order gradient terms in a continuum
    description of dislocation dynamics}, 2003 {\it Acta Mater.}
  {\bf 51} 1271--81
\bibitem{zaiser06} Zaiser M, {\it Scale invariance in plastic flow of
    crystalline solids}, 2006 {\it Adv. Phys.} {\bf 55} 185--245
\bibitem{miguel02} Miguel M C, Vespignani A, Zaiser M and Zapperi S,
  {\it Dislocation jamming and Andrade creep}, 2002 \PRL {\bf 89}
  165501
\bibitem{miguel05} Miguel M C, Moretti P, Zaiser M and Zapperi S, {\it
    Statistical dynamics of dislocations in simple models of plastic
    deformation: Phase transitions and related phenomena}, 2005 {\it
    Mat. Sci. Eng.} A {\bf 400--401} 191--8
\bibitem{miguel07} Miguel M C, Laurson L and Alava M, {\it Material yielding 
    and irreversible deformation mediated by dislocation
    motion}, 2008 {\it Eur. Phys. J.} B {\bf 64} 443--50
\bibitem{wilkens67} Wilkens M, {\it Das Spannungsfeld einer Anordnung 
    von regellos verteilten Versetzungen}, 1967 {\it Acta
    Met.} {\bf 15} 1412--7
\bibitem{wilkens69} Wilkens M, {\it Das mittlere Spannungsquadrat
    $\langle \sigma^{2} \rangle$ begrenzt regellos verteilter
    Versetzungen in einem zylinderf\"ormigen K\"orper}, 1969
  {\it Acta Met.} {\bf 17} 1155--9 
\bibitem{zaiser01} Zaiser M, Miguel M-C and Groma I, {\it Statistical
    dynamics of dislocation systems: The influence of
    dislocation--dislocation correlations}, 2001 \PR B {\bf 64} 224102
\bibitem{zaiser02} Zaiser M and Seeger A, {\it Long-range internal
    stresses, dislocation patterning and work hardening in crystal
    plasticity}, 2002, {\it
    Dislocations in Solids Vol. 11}, ed F R N Nabarro and
  M S Duesbery (Elsevier)
\bibitem{csikor04} Csikor F F and Groma I, {\it Probability
    distribution of internal stress in relaxed dislocation systems},
  2004 \PR B {\bf 70} 064106
\bibitem{krivoglaz69} Krivoglaz M A 1969 {\it Theory of x-ray and
    thermal neutron scattering by real crystals} (New York:
  Plenum)
\bibitem{csikor05} Csikor F F, Kocsis B, Bak\'o B and Groma I,
  {\it Numerical characterisation of the relaxation of dislocation
    systems}, 2005 {\it Mat. Sci. Eng.} A {\bf 400--401} 214--7
\bibitem{csikor06} Csikor F F and Zaiser M, {\it Scaling and glassy
    dynamics in the relaxation of dislocation systems}, 2006 {\it
    Proc. Int. Conf. on Statistical Mechanics of Plasticity and
    Related Instabilities (29 August--2 September 2005 Bangalore)} ed
  M Zaiser {\it et al} (Proceedings of Science) 058
\bibitem{hirthlothe82} Hirth J P and Lothe J 1982 {\it Theory of
    Dislocations} (New York: Wiley-Interscience)
\bibitem{bako06} Bak\'o B, Groma I, Gy\"orgyi G and Zim\'anyi G, {\it
    Dislocation patterning: The role of climb in meso-scale
    simulations}, 2006 {\it Comp. Mater. Sci.} {\bf 38} 22--8
\bibitem{ovchinnikov78} Ovchinnikov A A and Zeldovich Ya B, {\it Role
    of density fluctuations in bimolecular reaction kinetics}, 1978
  {\it Chem. Phys.} {\bf 28} 215--8
\bibitem{groma06} Groma I, Gy\"orgyi G and Kocsis B, {\it Debye
    screening of dislocations}, 2006 \PRL {\bf 96} 165503
\bibitem{thomson06} Thomson R, Koslowski M and LeSar R, {\it
    Energetics and noise in dislocation patterning}, 2006
  \PR B {\bf 73} 024104 
\bibitem{nabarro00} Nabarro F R N, {\it Complementary models of
    dislocation patterning}, 2000 {\it Phil. Mag.} A {\bf 80} 759--64
  \endbib
\end{document}